**Noise-tolerant correlated coincidence imaging based on super-correlated light at 1550 nm**

Yu Yan[1], Jiamin Li[1,2*], Ruikang Li[1], Yanqiang Guo[1,2,3], Jiang Qiu[1], Shuangping Han[1,2], Zihua Liu[1,2], Jianyong Hu[1,4], Chengbing Qin[1,4*] and Liantuan Xiao[1,2,3,4*]

[1] College of Physics and Optoelectronic Engineering, Taiyuan University of Technology, Taiyuan, Shanxi 030024, China.

[2] Shanxi Key Laboratory of Precision Measurement Physics, Taiyuan University of Technology, Taiyuan, Shanxi 030024, China.

[3] Key Laboratory of Advanced Transducers and Intelligent Control System, Ministry of Education, Taiyuan University of Technology, Taiyuan, Shanxi 030024, China.

[4] Institute of Laser Spectroscopy, State Key Laboratory of Quantum Optics Technologies and Devices, Shanxi University, Taiyuan, Shanxi 030006, China.

*: Corresponding authors, email:

Jiamin Li: lijiamin@tyut.edu.cn

Chengbing Qin: chbqin@sxu.edu.cn

Liantuan Xiao: xlt@sxu.edu.cn

**Abstract**

Single-photon-level imaging at 1550 nm is a key driver for significant advancements in the next-generation laser detection technology. This cutting-edge approach plays a vital role in space ranging, target recognition, and three-dimensional remote sensing. However, it has faced severe challenges such as insufficient noise-tolerant performance. Here, we introduced noise-tolerant correlated coincidence imaging (CCI) based on super-correlated light. The light source, generated through nonlinear interaction between a pulsed laser and a photonic crystal fiber, exhibits a broader power-law photon number probability distribution and extremely strong photon correlation (with second-order correlation function $g^{(2)}(0)$ up to 18,166). Our noise-tolerant CCI can resist random environmental noise up to 100,000 times stronger than the echo signal photons. Super-correlated light offers an exceptionally strong noise tolerance for single-photon-level imaging in extreme environments with intense noise, paving the way for the future development of extremely sensitive light detection.

**Introduction**

Active imaging technology operating at a wavelength of 1550 nm combines the dual advantages of low transmission loss and human eye safety (*1-3*), serving as a core driver in advancing frontier fields such as target recognition, biomedical imaging, and safety monitoring (*4-7*). High-power lasers are commonly used for target object illumination in conventional imaging systems, while traditional photodetectors are employed to capture echo signals (*8, 9*). Single-photon-level imaging technology overcomes the limitations of traditional systems by enabling robust detection under illumination conditions that are far below the threshold of conventional imaging systems (*10, 11*).

Despite single-photon imaging technology having achieved outstanding performance, it still faces a frontier question: noise interference (*12, 13*). The simultaneous influence of environmental noise and dark counts from detectors significantly compromises both the signal-to-noise ratio and image contrast, resulting in the masking of accurate target information by substantial noise. Currently, optical modulation and computing algorithms are employed as an effective strategy to suppress a large amount of noise interference (*14-19*). Recent advances in correlated coincidence imaging have proposed novel schemes based on exploiting non-classical light sources, significantly enhancing image quality (*20-28*). Key properties enabling these improvements include photon correlations in entangled photon pairs and bright squeezed light. For instance, Zhang demonstrated that correlated coincidence imaging, achieved through spatial correlation between photon pairs, exhibits a 25-fold greater noise tolerance than conventional imaging (*22*). Similarly, in ref (*23*), spatial-momentum entanglement is employed to extract biphoton coincidence counts in correlated coincidence imaging using a covariance algorithm, effectively eliminating uncorrelated noise. This method is capable of withstanding environmental noise up to

155 times stronger than the classical signal. Notably, this inherent photon correlation property provides a natural advantage in suppressing environmental noise. Consequently, developing novel types of light sources with strong correlations is a key strategy for fundamentally resisting noise.

Strong photon correlation can not only be achieved through the entanglement process with phase matching, but also via the generation of intense photon number fluctuations. Recent studies have demonstrated that the bright squeezed vacuum is a nonclassical light source exhibiting strong photon correlations ($g^{(2)}(0)$=3) ($29$). This strong correlation arises from significant photon number fluctuations, which result in nonclassical statistical behavior that deviates markedly from the Poisson distribution, giving rise to a substantially higher probability of multiphoton emission ($30, 31$). In general, enhancing photon number fluctuation and multiphoton emission probability have been achieved in both linear and nonlinear optical systems. In linear systems, modulation of photon number probability distribution and photon correlations can be achieved through optical scattering systems (such as rotating ground glasses) or optical modulators ($32, 33$). On the other hand, strong photon correlation properties have been observed in various nonlinear optical systems, including those involving nonlinear optical crystals ($34$), cold atom systems ($35, 36$), quantum dots ($37$-$40$), and two-dimensional materials ($41$). Furthermore, studies demonstrate that the nonlinear interaction between bright squeezed vacuum and single-mode fiber generates enhanced photon correlation, enhancing the correlation properties of the light field. The second-order correlation function $g^{(2)}(0)$ has reached the current maximum reported value of 170 ($42$).

In this work, we propose a noise-tolerant correlated coincidence imaging (CCI) system based on a super-correlated light source at 1550 nm. The super-correlated light sources that exhibit unprecedented photon correlations are achieved through complex nonlinear effects between pulsed lasers and engineered photonic crystal fibers. The measured intensity correlation function $g^{(2)}(0)$ reached 18,166. This super correlation arises from the utmost multiphoton bundled emissions and a broader photon number distribution than classical light. Our imaging scheme utilizes biphoton coincidence detection from the source's dual output paths. Remarkably, the system maintains imaging capability under environmental noise 100,000 times stronger than signal photons. The results demonstrated that CCI can resist random environmental noise up to 100,000 times stronger than signal photons. Compared to the most advanced correlated coincidence imaging schemes using other nonclassical light sources, the noise-tolerant ability of our CCI scheme has improved by 3 orders of magnitude (*20-23*). When the noise photon counts reach 50,000 and 20,000 times that of the signal photons respectively, the peak signal-to-noise ratio (PSNR=13.68 dB) and contrast-to-noise ratio (CNR=4.46) of our CCI are comparable to those of traditional photon-counting imaging (PCI) without external noise. Super-correlated light sources exhibit stronger photon correlations, which translates to unparalleled noise tolerance, thereby serving as a novel light source for single-photon-level correlated coincidence imaging.

**Results**

**Principle of noise-tolerant correlated coincidence imaging**

We introduced a noise-tolerant correlated coincidence imaging (CCI) technique based on pulsed super-correlated light sources (SCL) with high $g^{(2)}(0)$. For CCI, one path of the SCL is used to illuminated the target object, and scattered light from the object is used as the detection light (CH1, Fig. 1A). The other path serves as reference light (CH2), remaining locally, which is combined with detection photons for coincidence measurements. The arrival pattern of photons is shown in Fig. 1A, where the red and blue balls represent the correlated photons and noise photons, respectively. The arrival times of noise photons are distributed randomly across all pulses. In contrast, correlated photons from a single pulse arrive simultaneously within their pulse width. Therefore, coincidence counts from super-correlated light can only be measured if both photons arrive within the coincidence window $t_1$ ($t_1 \ll t_2$, where $t_2$ is the pulse period) of the same pulse. Noise photons can be effectively suppressed through coincidence measurements between the detection and reference paths with a limited time window.

To quantify correlated coincidence imaging quality, we can determine that the coincidence count per second of the object area $C_o$ is $C_o = (g^{(2)}(0) \cdot N_r N_s + N_r \dfrac{t_1}{t_2} N_n) \times t_1$, and that of the background area $C_b$ is $C_b = N_r \dfrac{t_1}{t_2} N_n t_1$, where $N_r$, $N_s$, and $N_n$ denote the reference, signal and noise photon count per second, respectively (see derivation in note S1). Consequently, the difference between the coincidence counts per second of the object area and the background area can be expressed as $C_o - C_b = g^{(2)}(0) \cdot N_r N_s t_1$, exhibiting a positive correlation with $g^{(2)}(0)$. Under identical conditions, higher $g^{(2)}(0)$ values lead to greater differences, thereby enhancing imaging contrast.

Here, we utilized the contrast-to-noise ratio (CNR) to evaluate the noise tolerance of CCI based on super-correlated light sources with high $g^{(2)}(0)$ in an intense noise environment. The CNR of the reconstructed image can be expressed as (note S1):

$$\mathrm{CNR} = \frac{\left| C_\mathrm{o} - C_\mathrm{b} \right|}{\sqrt{\sigma_{C_\mathrm{o}}^2 + \sigma_{C_\mathrm{b}}^2}}$$

$$= \frac{g^{(2)}(0) N_r N_s}{\sqrt{(g^{(2)}(0))^2 \cdot \left( N_r N_s \right)^2 \left( \dfrac{1}{N_r} + \dfrac{1}{N_s} \right) + 2\left( N_r \dfrac{t_1}{t_2} N_n \right)^2 \left( \dfrac{1}{N_r} + \dfrac{t_2}{t_1 \times N_n} \right)}}, \quad (1)$$

where $\sigma_{C_\mathrm{o}}$ and $\sigma_{C_\mathrm{b}}$ represent the standard deviations of the coincidences of the object and background areas. For comparison, we also present the results of conventional single-photon imaging (*i.e.* photon-counting imaging (PCI)), where the value of each pixel corresponds to the number of detected photons. The corresponding CNR calculation method is detailed in note S1. To quantitatively analyze the influence of $g^{(2)}(0)$ value and noise photons on the imaging CNR, we assume that $N_r = N_s$, and define the ratio of total photons to signal photons, $R$, as follows:

$$R = \frac{N_\mathrm{n} + N_\mathrm{s}}{N_\mathrm{s}}. \quad (2)$$

The simulation results of the CNR of CCI and PCI are shown in Figs. 1B and 1C. For CCI, where $g^{(2)}(0)$ remains constant with $R$ is gradually increasing from $10^0$, CNR maintains good stability across a wide range of $R$ values. Only when $R$ reaches significantly high values does the CNR begin to decline gradually. Comparing the simulations with different photon correlation, we find that a high $g^{(2)}(0)$ value has a strong resistance to intense noise. For instance, when $g^{(2)}(0)=10^4$, CNR exhibits excellent robustness across the range of $R=10^0$ to $10^4$. As $R$ increases to $10^5$, CNR gradually degrades. In contrast, the dotted line represents the traditional PCI, which exhibits significantly degraded CNR as $R$ increases. Figure 1C shows the CNR variation as a function of $g^{(2)}(0)$ values across different $R$ conditions. The CNR exhibits progressive enhancement with increasing $g^{(2)}(0)$. Notably, at $R=10^5$ and $g^{(2)}(0)=10^4$, the CNR can reach 9.4, demonstrating that the super-correlated light source with a high

$g^{(2)}(0)$ can significantly enhance image quality. Therefore, super-correlated light sources with high $g^{(2)}(0)$ exhibit excellent noise resistance in CCI. Consequently, developing a light source with a high $g^{(2)}(0)$ value is crucial for advancing noise-resistant correlation imaging.

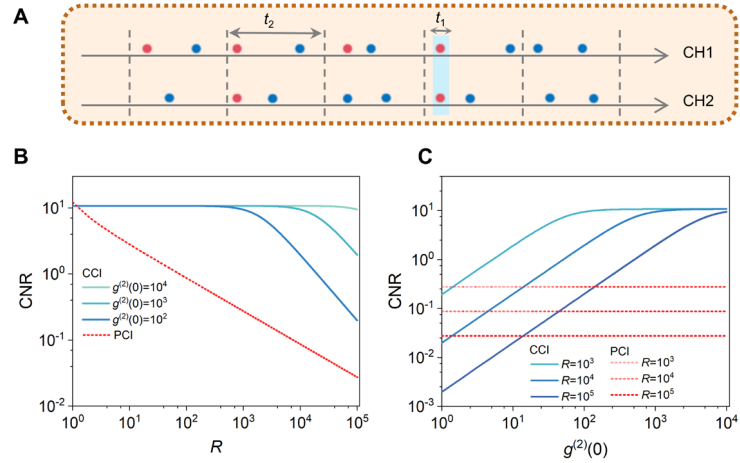

**Fig. 1 Principle of noise-tolerant correlated coincidence imaging by super-correlated light.** (**A**) Schematic diagram of correlated coincidence imaging through the coincidence counts from the dual channels. Red and blue balls represent the photons detected from the super-correlated light source and noise. (**B**) The simulated CNR for CCI and PCI as a function of R across $g^{(2)}(0)$. (**C**) The simulated CNR as a function of $g^{(2)}(0)$ across different $R$.

**Generation and characterization of super-correlated light**

To develop a nonclassical light source with giant $g^{(2)}(0)$ vales for correlated coincidence imaging applications, we proposed the generation of super-correlated light through nonlinear interaction between a pulse laser and optimized photonic crystal fiber (PCF) (*42-45*). The details have been reported in our recent work (*31*).

We first generated a super-correlated light at 1550 nm by using a picosecond pulse laser at 1064 nm to pump an optimized photonic crystal fiber. Additional details are provided in Methods and Supplementary Information (notes S3, figs. S2 and S3). Subsequent characterization revealed the unique properties of this super-correlated light source. The pump power can significantly alter the dominant nonlinear effects, thereby altering both the photon emission rate and the second-order correlation function $g^{(2)}(0)$ of the super-correlated light (*46, 47*). When the power of the pump pulse reaches the generation threshold of 190 mW for the 1550 nm spectrum (inset of Fig. 2A), a large number of photons emits from the PCF, leading to a sharp increase in the photon count rates, as shown in Fig. 2A. As the pump power exceeds 340 mW, the growth of photon counts slows down due to the gradual emergence of other spectral components. To confirm its super-correlation characteristics, we measured the second-order correlation function $g^{(2)}(0)$ of the generated super-correlated light source based on the HBT system (Fig. 2B, the HBT device is shown in fig. S2). The $g^{(2)}(0)$ value increases sharply as the pump power approaches the threshold of super-correlated light, which can be considered as the proportion of generated correlation photons increases significantly. With a further increase in pump power, the $g^{(2)}(0)$ value shows a decreasing trend, due to a large number of uncorrelated photons generated as the gain continues to increase. When the pump power exceeds 450 mW, $g^{(2)}(0)$ gradually decreases to 1, attributed to the increasing dominance of coherent and thermal light components. It is in good agreement with the theoretical prediction (fig. S1). Therefore, the maximum $g^{(2)}(0)$ can be determined under optimal conditions. For example, at a pump power of 260 mW, $g^{(2)}(0)$ reaches its maximum, with a value of 18166 (Fig. 2C). We further characterized the $g^{(2)}(0)$ at different mean photon numbers $\langle n \rangle$ from $3.9 \times 10^{-3}$ to $6.1 \times 10^{-5}$, achieved by attenuation with a neutral density filter (Fig. 2D). The measured $g^{(2)}(0)$ values exhibit a significant increase from

210 to 10738. Contrary to the traditional understanding that second-order correlations are insensitive to the variation of $\langle n \rangle$, $g^{(2)}(0)$ of the super-correlated light increases significantly as the $\langle n \rangle$ decreases. This unusual behavior is attributed to the presence of a large number of multiphoton pulses in the super-correlated light. As the attenuation ratio increases, single- and two-photon pulses ($N \leqslant 2$) are suppressed, resulting in an increase in photon sparsity in the time domain, which leads to a higher proportion of multiphoton pulses. This increase, in turn, significantly enhances $g^{(2)}(0)$ (note S4 and fig. S4). Compared to coherent and thermal light, the $g^{(2)}(0)$ of the super-correlated light is improved by at least 4 orders of magnitude (*37, 48, 49*). Furthermore, we measured the coincidence counts per second and $g^{(2)}(0)$ over varying measurement times while maintaining a constant pump power of 260 mW. As measurement time increases, the deviations of coincidence counts per second and $g^{(2)}(0)$ are significantly reduced (Fig. 2E). This phenomenon clearly indicates that the stability of the photon correlation of the light source varies over time. In practical applications, the measurement time needs to be optimized to achieve the required stability.

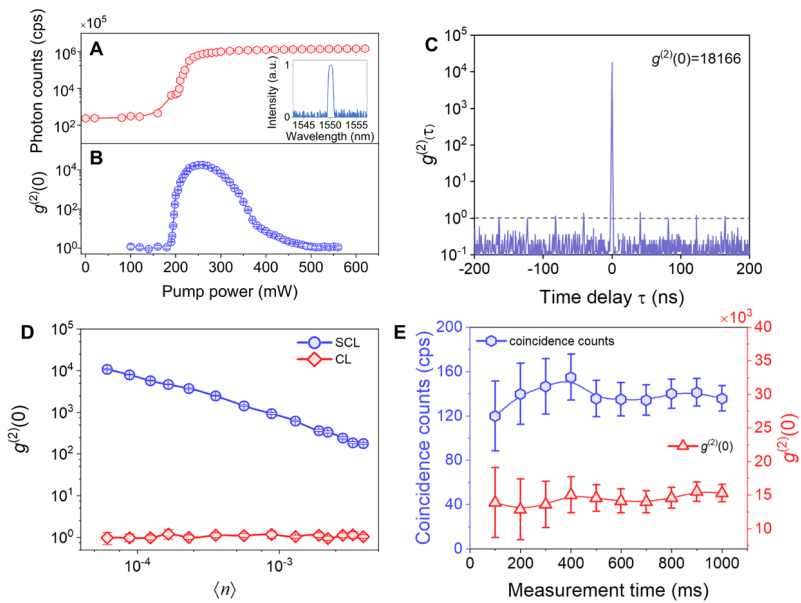

**Fig. 2 Schematic diagram of the experimental setup and characterization of super-**

**correlated features.** (**A**) Photon counts per second of SCL as a function of pump power. The inset shows the filtered spectrum centered at 1550 nm. (**B**) Measured $g^{(2)}(0)$ as a function of pump power. (**C**) Second-order correlation function $g^{(2)}(\tau)$ as a function of the time delay $\tau$, observed bunching and normalization peaks at $\tau=NT$, where $T=1/f=$ 41.06 ns and $N$ are integer values. (**D**) Comparison of $g^{(2)}(0)$ as a function of the mean photon number per pulse $\langle n \rangle$ for SCL and coherent light (CL). (**E**) The coincidence counts (coin-counts) and second-order correlation function $g^{(2)}(0)$ corresponding to different measurement times.

To verify the multiphoton characteristics of the super-correlated light source, the photon number probability distribution was measured and compared with that of coherent light and thermal light, which satisfied the Poisson and Bose-Einstein distributions, respectively (notes S5 and S6, fig. S5). Figure 3A shows the photon number probability distribution at different pump powers. As expected, the mean photon number and the photon emission probability increased intuitively with the increase in the pump power. Super-correlated light exhibits a broader photon number probability distribution compared with coherent light and thermal light, revealing a significant enhancement of multiphoton events.

To quantify the deviation of the multiphoton emission probability, we define the ratio $\xi_{\mathrm{SCL/ML}}$ as:

$$\xi_{\mathrm{SCL/ML}} = \frac{P_{\mathrm{SCL}}(n)}{P_{\mathrm{ML}}(n)} \qquad (3)$$

where $P_{\mathrm{SCL}}(n)$ and $P_{\mathrm{ML}}(n)$ represent the $N$-photon bundled emission probabilities of super-correlated light and other light (coherent light as $P_{\mathrm{CL}}(n)$ and thermal light as $P_{\mathrm{TL}}(n)$), respectively. For instance, at the pump powers of 230 mW, 260 mW, and 290 mW (corresponding to $\langle n \rangle = 1.1 \times 10^{-2}$, $4.4 \times 10^{-2}$, and $1.3 \times 10^{-1}$), the multiphoton emission probabilities are all significantly enhanced relative to those of coherent light and thermal light. In particular, the emission probabilities for 16 photons increased by 36, 29 and 23 orders of magnitude relative to coherent light (Fig. 3B) and by 23, 16 and 10 orders of magnitude relative to thermal light (Fig. 3C), respectively. In other words, the lower pump powers the smaller mean photon numbers and the greater the enhancement of multiphoton events. These results demonstrate a large number of extreme multiphoton events, which significantly deviate from both the Poisson distribution and the Bose-Einstein distribution. We also confirm that these enhancements can be further expanded by attenuating the mean photon number $\langle n \rangle$ while holding the pumping power. As shown in Fig. 3D, the pump power is maintained at 260 mW, as the mean photon number drops to $\langle n \rangle = 2.8 \times 10^{-4}$, the probability of detecting 16 photons reaches as high as $4.9 \times 10^{-10}$, representing the enhancement of 60 and 47 orders of magnitude compared to coherent light and thermal light, respectively (Figs. 3E and 3F). Therefore, as the mean photon number $\langle n \rangle$ decreases, single- and two-photon pulses ($N \leqslant 2$) are annihilated, while the probability of multiphoton emission increases relatively. This result is intrinsically related to the increase of $g^{(2)}(0)$ with the decrease of $\langle n \rangle$, as shown in Fig. 3G. These super-correlation properties and extreme multiphoton events offer a novel approach to noise-tolerant correlated coincidence imaging under extremely weak light conditions, where weaker light intensity corresponds to higher photon correlation.

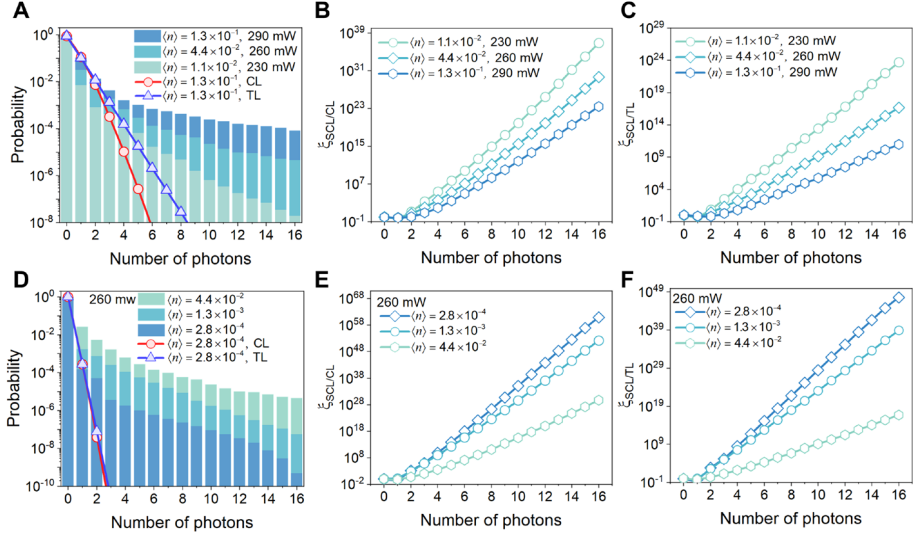

**Fig. 3 Measurement of photon number probability distribution of the generated super-correlated light.** (**A**) Photon emission probability of 1550 nm under different pump powers with fixed attenuation. The triangle and circle symbols represent the simulated photon number probability distributions at $\langle n \rangle = 1.3 \times 10^{-1}$ for TL and CL, respectively. The ratio of the multiphoton emission probabilities of SCL to CL (**B**) and TL (**C**) under different pump powers. (**D**) Photon emission probability for varying mean photon numbers per pulse $\langle n \rangle$ at the same pump power. The triangle and circle symbols represent the simulated photon number probability distributions at $\langle n \rangle = 2.8 \times 10^{-4}$ for TL and CL, respectively. The ratio of the multiphoton emission probabilities of SCL to CL (**E**), and TL (**F**) under the same pump power and different $\langle n \rangle$ values.

**Noise-tolerant correlated coincidence imaging based on the super-correlated light**

Note that the super-correlated light reaches its maximum $g^{(2)}(0)$ value at a pump power of 260 mW (Fig. 2C). Therefore, we performed noise-tolerant correlated coincidence imaging under this pump power condition. The imaging scheme is shown in Fig. 4A.

The super-correlated light source is divided into two beams by the fiber beam splitter (FBS1). One beam is stored locally as the reference light, which is free from external noise and connected to the superconducting nanowire single-photon detector (SNSPD), with its mean photon number adjustable via a variable optical attenuator (VOA). Another beam forms the detection path, being collimated by a collimator (COL) and then incident on a dual-axis galvanometer scanner (2-axis galvo) via a 45° perforating mirror (PERM) to scan the target object. The scattered photons from the target object are collected by the lens and coupled to the high-transmission port (99%) of the fiber beam splitter (FBS2). Environmental noise photons are injected through the low-transmission port (1%) of the FBS2, where it is mixed with the detected signal photons and transmitted to the SNSPD. For comparison, we evaluated the results of PCI under the same noise conditions.

To evaluate the distinguishability of the target object, the reconstructed image obtained with high photon counts (under noise-free conditions) and long measurement time serves as the true (reference) image, as shown in fig. S6C, compared this reference imaging with CCI and PCI under different noise conditions ($R$ = 1, 100, 1000, and 100,000) to calculate PSNR and CNR. The calculation workflows for PSNR and CNR are shown in "Methods". The color bars represent the coincidence counts and photon counts per pixel, respectively. Under noise-free conditions (*i.e.*, $R$ = 1), CCI achieves better imaging performance (PSNR = 28.32 dB, CNR = 8.02) than PCI (PSNR = 13.68 dB, CNR = 4.46), allowing for more distinct target identification. Remarkably, even with $R$ increasing by three orders of magnitude (from 1 to 1000), the target object and background regions can still be well distinguished for CCI (Figs. 4B-4D). For instance, at $R$ = 1000, the system exhibits excellent robustness, with a PSNR of 24.91 dB and a

CNR of 6.14. Until $R = 100000$, there is partial blurring between the target object and the background area (Fig. 4E). During the gradual increase in noise photons, both PSNR and CNR showed a slow decline (Figs. 4J and 4K). The noise tolerance limit is reached when the accidental coincidence counts from the background area become comparable to the coincidence counts of the object area. However, for PCI, as the $R$ values increased, the signal photon of the target object was rapidly submerged in the background noise (Figs. 4F-4I), resulting in dramatic reductions in both PSNR and CNR (Figs. 4J and 4K). Furthermore, as shown in Figs. 4J and 4K, CCI achieve PSNR and CNR values comparable to noise-free PCI even at extremely high $R$ of 50,000 and 20,000 respectively. This result demonstrates the exceptional robustness of the super-correlated light source, as its strong photon correlation properties enable high-quality correlated coincidence imaging even in intense noise environments.

To elucidate the noise-tolerant mechanism of CCI over PCI, we performed an analysis by sampling 16 pixels from both object and background areas. For each area, we systematically evaluated the visibility ($V = \dfrac{|I_1 - I_2|}{|I_1 + I_2|}$) of average coincidence counts and photon counts per pixel at different $R$ values (Fig. 4L). Here, $I_1$ represents the average coincidence counts per pixel or average photon counts per pixel of the object area, $I_2$ represents the average coincidence counts per pixel or average photon counts per pixel of the background area. For PCI, when $R$ exceeds 100, the PCI visibility drops below 0.1, demonstrating that signal photons are overwhelmed by noise photons. In contrast, the visibility of CCI is 0.99. Even when the $R$ increases to 100,000, it remains 0.24. Thus, CCI achieves superior noise tolerance relative to PCI. This pronounced advantage establishes a promising framework for developing robust optical systems capable of maintaining a high signal-to-noise ratio in single-photon level imaging under extreme environments.

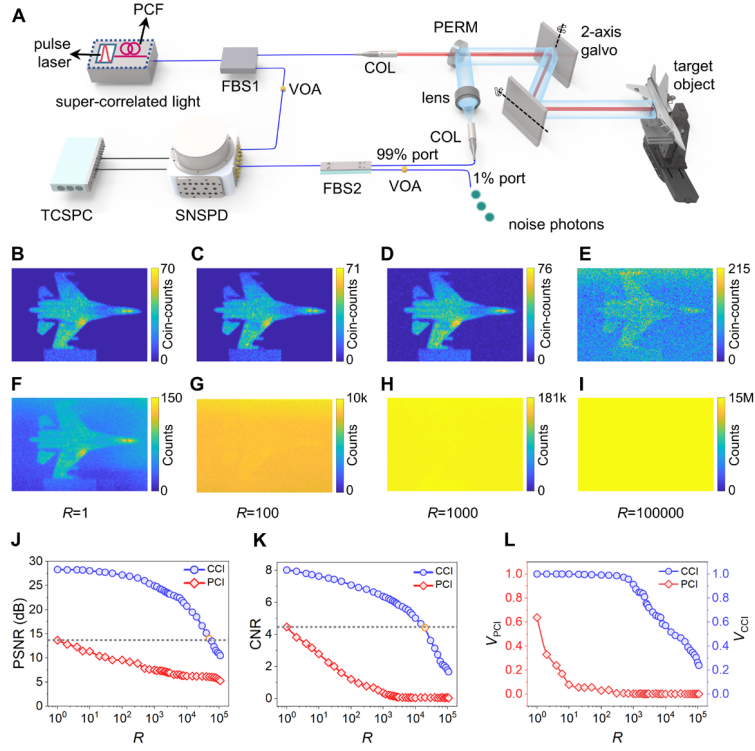

**Fig. 4 Experiments of correlated coincidence imaging and analysis. (A)** Schematic diagram of the imaging system. FBS, fiber beam splitter; VOA, variable optical attenuator; COL, collimator; PERM, 45° perforating mirror; 2-axis galvo, dual-axis galvanometer scanner. (**B-E**) Reconstructed images of CCI at different $R$, ranging from 1-100000. The color bar indicates the coincidence count (coin-counts) values. (**F-I**) Reconstructed images of PCI at different $R$, ranging from 1-100,000. The color bar indicates the photon count values. (**J and K**) The PSNR and CNR of CCI and PCI under different $R$. (**L**) The visibility ($V$) of average coincidence counts per pixel of CCI and average photon counts per pixel of PCI in object and background areas, respectively.

## Discussion

We demonstrate noise-tolerant correlated coincidence imaging based on super-correlated light sources, aiming to enhance noise tolerance in extreme environments by leveraging the unique properties of these light sources. The environmental noise is suppressed by detecting the coincidence counts of correlated photons with time correlation (*50*). Compared with PCI, coincidence count rate and anti-noise performance have been significantly improved, due to its extremely high $g^{(2)}(0)$ value. These improvements can dramatically enhance the imaging quality in practical applications (*51, 52*). It offers a novel solution for single-photon-level imaging in complex environments. Next, for tiny mean number of photons per pulse $\langle n \rangle$, the exceptionally high probability of *N*-photon emission by the super-correlated light source enables us to image the target by statistically counting the coincidences between multiple channels, even if the echo pulses are extremely weak. This characteristic may be helpful for quantum imaging in environments where the signal photons are significantly attenuated but with intense background light and noise (*53*). Finally, another advantage of the super-correlated light source is its full-fiberization, which can be integrated into subsequent work and offers stable environmental robustness. This advantage significantly reduces system complexity, making it highly valuable for further expanding the applications of super-correlated light sources.

In summary, we have developed a super-correlated light source and demonstrated its capability for high-contrast imaging under extreme noise conditions. By leveraging the nonlinear interaction between pulsed lasers and optimized PCF, super-correlated light sources with powerful second-order correlation functions and extreme multi-photon events have be generated. The $g^{(2)}(0)$ is as high as 18,166, which is 2 orders of magnitude higher than the current optimal report. When the mean photon number is $\langle n \rangle = 2.8 \times 10^{-4}$, the emission probability of 16 photons is increased by 60 and 47 orders of magnitude, respectively, compared with coherent light and thermal light. At the single-photon level, correlated coincidence imaging based on strong photon correlation can resist random environmental noise exceeding 100,000 times, which is 3 orders of magnitude more robust than the most advanced nonclassical light source correlation imaging schemes (fig. S7). When the noise photons exceeded signal photons by 50,000 and 20,000-fold, respectively, the PSNR (13.68 dB) and CNR (4.46) of CCI reached the levels of PCI without external noise interference, indicating its excellent robustness. Each pixel of the reconstructed image is generated at the single-photon level, which enables the target to be identified even in extremely weak light environments. Moreover, by adjusting the pump power and filtering range, super-correlated light across the C+L band can be generated. This capability offers new solutions for parallel single-photon LiDAR systems and quantum random number generation (54, 55), highlighting the versatility and potential of super-correlated light in advanced photonic technologies. Notably, when the biphoton coincidence counts reach the noise tolerance limit, the super-correlated light source can leverage its inherent multiphoton emission probability to construct higher-order correlations, achieving superior noise resistance. Optical imaging systems based on super-correlated light have opened up new opportunities for target recognition in extreme environments.

**Materials and Methods**
**Experimental design**

The schematic diagram of the super-correlated light-generating device is shown in fig. S2. The seed light is a mode-locked laser based on semiconductor saturable absorption mirror (SESAM, SAM-1064-21-3 ps, BATOP). The laser system comprises the following key components: a SESAM, a semiconductor laser (LD1, VLSS-976-B-600-FA, CONNET), a wavelength division multiplexer (WDM, MCWDM-9764-00-H1-10-

L-FA, DK Photonics), a fiber Bragg grating (FBG, MCFBG-1064-300-H1-0.3-10-FA, DK Photonics), a fiber isolator (ISO, MCI-1064-00-P-S-H1-10-N-FA, DK photonics), and a 0.6 m long single-mode ytterbium-doped fiber (YDF, Yb/1200, LIEKKI) as the gain medium. The LD1 has a central wavelength of 976 nm and a maximum output power of 600 mW. The SESAM exhibits a modulation depth of 14% and a reflectance of 78% at 1064 nm. The FBG has a reflectance of 10% at 1064 nm and a spectral bandwidth of 0.3 nm. The ISO1 features an isolation bandwidth of ±5 nm and an isolation ratio of 30 dB. When the pumping power of LD1 reaches 120 mW, the seed laser achieves mode-locking, with an output center wavelength of 1064 nm, a pulse width of approximately 100 picoseconds, a repetition rate of 24.35 MHz, and an average power of 35 mW. Then the power of the pulsed laser is amplified. It consists of a semiconductor laser (LD2, VLSM-976-B-5, CONNET) with a maximum output power of 5 W, a fiber combiner (FC, MCMPL-976/20-105/125/22/H1-1064/10-10/125D-08-2, DK Photonics), a 3 m long double-clad ytterbium-doped fiber (DCYDF, LMA-YDF-10/125, Nufern) and a fiber isolators (ISO2, MCHI-1064-30-P-50-S-10/125DC-10-B-N, DK Photonics), all of which combine to form a pulsed laser with an average power of 1 W. The output pigtail of the pulsed laser is coupled to a mode field adapter (MFA, MCMFA-1064-B-20-10/125DC-H1-08-P2, DK Photonics) to convert the mode field area, and then fused with the photonic crystal fiber (PCF, SC-PRO, YSL Photonics). The output light is collimated by an optical collimator (COL, MCCOL-1550-50-2-10-S2-08-L-S, DK Photonics) and then passed sequentially through an acousto-optic tunable filter (AOTF, AOTF-PRO2, YSL Photonics), neutral density filters (Att, COFV2-40M, JCOPTIX). The filtered signal is subsequently coupled into a dense wavelength division multiplexer (DWDM, MCDWDM-1-C34-

00-S2-08-L-FA, DK Photonics) operating at a center wavelength of 1550 nm with a channel bandwidth of 0.8 nm. The spectrum is measured using a fiber optic spectrometer (AQ6370, Yokogawa). The fiber beam splitter (FBS, MCFBT-12-1550-50-00-16-S2-08-L-FA, DK Photonics) evenly splits the light and is connected to a 16-channel superconducting nanowire single-photon detector (SNSPD, P-SPD-16S, PHOTEC). The detected signal is recorded and processed using time-correlated single-photon counting (TCSPC, MT16, SIMINICS). Based on the recorded photon time series, photons detected by the same pulse at the same time are counted as coincidence counts. In contrast, photons detected by adjacent pulses are counted as accidental coincidence counts. The multi-channel photon count rates and coincidence count rates obtained from the measurements are used to calculate the photon number probability distribution and the second-order correlation function $g^{(2)}(0)$.

**Detection of the photon number probability distribution**

In this experiment, each photon is detected by the SNSPD, and the arrival time of the photons is recorded via TCSPC. The multi-body coincidence function in TCSPC enables the measurement of simultaneous multi-photon events detected within a defined time window. First, the photon arrival times of all SNSPD channels were aligned by TCSPC, and then all the detected multi-photon events were classified and counted. In this work, the repetition frequency of 24.35 MHz corresponds to a time window of 41.06 ns.

**Noise-tolerant correlated coincidence imaging based on a super-correlated light**

Noise-noise correlated coincidence imaging is performed at a pump power of 260 mW,

resulting in a relatively high second-order correlation function, $g^{(2)}(0)$. The reference path and the detection path are separated by FBS1 (MCFBT-12-1550-50-00-2-S2-08-L-FA, DK Photonics). A 45° perforating mirror (PERM) and dual-axis galvanometer scanner (2-axis galvo, GVS012/M, Thorlabs) are used to transmit signal light and scan target objects. The signal photons and noise photons are mixed by FBS2 (MCFBT-12-1550-01-00-2-S2-08-L-FA, DK Photonics) and transmitted to SNSPD. A white LED (M1550L4, Thorlabs) powered by an LED driver (DC2200, Thorlabs) was used to generate random noise photons during imaging. The number of noise photons can be adjusted via a variable optical attenuator (VOA, MCVOA-1550-00-S2-10-L-FA, DK Photonics). To achieve higher image quality, the measurement time for each pixel is 500 ms (note S7 and fig. S8).

**Peak signal-to-noise ratio and contrast-to-noise ratio**

Peak signal-to-noise ratio (PSNR) and contrast-to-noise ratio (CNR) are used to quantify the imaging results. It can be described as:

$$\text{PSNR} = 10 \log_{10}(\frac{(MAX_{pixel})^2}{MSE}), \tag{4}$$

where $MAX_{pixel}$ is the maximum pixel value, and $MSE$ is the mean square error.

$$\text{CNR} = \frac{|\mu_{\text{o}} - \mu_{\text{b}}|}{\sqrt{\sigma_{\text{o}}^2 + \sigma_{\text{b}}^2}}. \tag{5}$$

where $\mu_{\text{o}}$ and $\mu_{\text{b}}$ represent the mean values of the pixel values of the object area and the background area, respectively. For PCI imaging, it is the photon count value; For CCI imaging, it is in accordance with the count value. $\sigma_{\text{o}}$ and $\sigma_{\text{b}}$ represent the standard deviations of the pixel values of the object area and the background area, respectively. Detailed information is available in fig. S6A[1].

**References**

1.  Z.-P. Li, J.-T. Ye, X. Huang, P.-Y. Jiang, Y. Cao, Y. Hong, C. Yu, J. Zhang, Q. Zhang, C.-Z. Peng, F.-H. Xu, J.-W. Pan, Single-photon imaging over 200 km. *Optica* **8**, 344-349 (2021).

2.  Z.-P. Li, X. Huang, Y. Cao, B. Wang, Y.-H. Li, W.-J. Jin, C. Yu, J. Zhang, Q. Zhang, C.-Z. Peng, F.-H. Xu, J.-W. Pan, Single-photon computational 3D imaging at 45 km. *Photonics. Res.* **8**, 1532-1540 (2020).

3.  Y. Hong, S. Liu, Z.-P. Li, X. Huang, P. Jiang, Y. Xu, C. Wu, H. Zhou, Y.-C. Zhang, H.-L. Ren, Z.-H. Li, J. Jia, Q. Zhang, C. Li, F. Xu, J.-Y. Wang, J.-W. Pan, Airborne single-photon LiDAR towards a small-sized and low-power payload. *Optica* **11**, 612-618 (2024).

4.  D.-B. Lindell, M. O'Toole, G. Wetzstein, Single-photon 3D imaging with deep sensor fusion. *ACM Transactions on Graphics* **37**, 1-12 (2018).

5.  X. Zhang, K. Kwon, J. Henriksson, J. Luo, M.-C. Wu, A large-scale microelectromechanical-systems-based silicon photonics LiDAR. *Nature* **603**, 253-258 (2022).

6.  S.-S. Chan, A. Halimi, F. Zhu, I. Gyongy, R.-K. Henderson, R. Bowman, S. McLaughlin, G.-S. Buller, J. Leach, Long-range depth imaging using a single-photon detector array and non-local data fusion. *Sci. rep.* **9**, 8075 (2019).

7.  S.-F. Wang, L. Liu, Y. Fan, A.-M. El-Toni, M.-S. Alhoshan, D.-D. li, F. Zhang, In Vivo High-resolution Ratiometric Fluorescence Imaging of Inflammation Using NIR-II Nanoprobes with 1550 nm Emission. *Nano. Lett.* **19**, 2418-2427 (2019).

8.  R.-X. Chen, H.-W. Shu, B.-T. Shen, L. Chang, W.-Q. Xie, W.-C. Liao, Z.-H. Tao, B.-J. E., X.-J. Wang, Breaking the temporal and frequency congestion of LiDAR by parallel chaos. *Nat. Photonics* **17**, 306-314 (2023).

9.  C.-A. Casacio, L.-S. Madsen, A. Terrasson, M. Waleed, K. Barnscheidt, B. Hage, M.-A. Taylor, W.-P. Bowen, Quantum-enhanced nonlinear microscopy. *Nature* **594**, 201-206 (2021).

10. K. Song, Y.-X. Bian, D. Wang, R.-R. Li, K. Wu, H.-R. Liu, C.-B. Qin, J.-Y. Hu, L.-T. Xiao, Advances and Challenges of Single-Pixel Imaging Based on Deep Learning. *Laser. Photonics. Rev.* **19**, 1 (2024).

11. G.-N. Gol'tsman, O. Okunev, G. Chulkova, A. Lipatov, A. Semenov, K. Smirnov, B. Voronov, A. Dzardanov, C. Williams, R. Sobolewski, Picosecond superconducting single-photon optical detector. *Appl. Phys. Lett.* **79**, 705-707 (2001).

12. P.-S. Blakey, H. Liu, G. Papangelakis, Y. Zhang, Z.-M. Leger, M.-L. Iu, A.-S.

Helmy, Quantum and non-local effects offer over 40 dB noise resilience advantage towards quantum lidar. *Nat. Commun.* **13**, 5633 (2022).

13. H.-C. Li, K.-M. Zheng, R. Ge, L.-B. Zhang, L.-J. Zhang, W.-J. He, B. Zhang, M. Wu, B. Wang, M.-H. Mi, Y.-Q. Guan, J.-R. Tan, H. Wang, Q. Chen1, X.-C. Tu, Q.-Y. Zhao, X.-Q. Jia, J. Chen, L. Kang, Q. Chen, null, P.-H. Wu, Noise-tolerant LiDAR approaching the standard quantum-limited precision. *Light. Sci. Appl.* **14**, 138 (2025).

14. W. Fan, G. Qian, Y. Wang, C.-R. Xu, Z. Chen, X. Liu, W. Li, X. Liu, F. Liu, X. Xu, D.-W. Wang, V.-V. Yakovlev, Deep learning enhanced quantum holography with undetected photons. *PhotoniX* **5**, 54-66 (2024).

15. X. Li, Y. Li, Y. Zhou, J. Wu, Z. Zhao, J. Fan, F. Deng, Z. Wu, G. Xiao, J. He, Y. Zhang, G. Zhang, X. Hu, X. Chen, Y. Zhang, H. Qiao, H. Xie, Y. Li, H. Wang, L. Fang, Q. Dai, Real-time denoising enables high-sensitivity fluorescence time-lapse imaging beyond the shot-noise limit. *Nat. biotechnol.* **41**, 282-292 (2023).

16. W.-W. Chen, S.-J. Feng, W. Yin, Y.-X. Li, J.-M. Qian, Q. Chen, C. Zuo, Deep-learning-enabled temporally super-resolved multiplexed fringe projection profilometry: high-speed kHz 3D imaging with low-speed camera. *PhotoniX* **5**, 306-317 (2024).

17. X. Zhang, B.-W. Wang, S. Li, K.-Y. Liang, H.-T. Guan, Q. Chen, C. Zuo, Lensless imaging with a programmable Fresnel zone aperture. *Sci. Adv.* **11**, eadt3909 (2025).

18. D. Shin, A. Kirmani, V.-K. Goyal, J.-H. Shapiro, Photon-Efficient Computational 3-D and Reflectivity Imaging With Single-Photon Detectors. *IEEE. T. Comput. Imag.* **1**, 112-125 (2015).

19. J. Rapp, V.-K. Goyal, A Few Photons Among Many: Unmixing Signal and Noise for Photon-Efficient Active Imaging. *IEEE. T. Comput. Imag.* **3**, 445-459 (2017).

20. T. Gregory, P.-A. Moreau, E. Toninelli, M.-J. Padgett, Imaging through noise with quantum illumination. *Sci. Adv.* **6**, eaay2652 (2020).

21. H. Defienne, M. Reichert, J.-W. Fleischer, D. Faccio, Quantum image distillation. *Sci. Adv.* **5**, eaax0307 (2019).

22. Y.-D. Zhang, Z. He, X. Tong, D.-C. Garrett, Rui Cao, L.-H.-V. Wang, Quantum imaging of biological organisms through spatial and polarization entanglement. *Sci. Adv.* **10**, eadk1495 (2024).

23. Z. He, Y.-D. Zhang, X. Tong, L. Li, L.-H.-V. Wang, Quantum microscopy of cells at the Heisenberg limit. *Nat. Commun.* **14**, 2339-2346 (2023).

24. G. Brida, M. Genovese, R. Berchera, Experimental realization of sub-shot-noise quantum imaging. *Nat. Photonics.* **4**, 227-230 (2010).

25. N. Samantaray, I. Ruo-Berchera, A. Meda, M. Genovese, Realization of the first sub-shot-noise wide field microscope. *Light. Sci. Appl.* **6**, e17005 (2017).

26. Y. Israel, S. Rosen, Y. Silberberg, Supersensitive polarization microscopy using NOON states of light. *Phys. Rev. Lett.* **112**, 103604 (2014).

27. G.-B. Lemos, V. Borish, G.-D. Cole, S. Ramelow, R. Lapkiewicz, A. Zeilinger, Quantum imaging with undetected photons. *Nature* **512**, 409-412 (2014).

28. M. Sanz, U. Las Heras, J.-J. Garcí´a Ripoll, E. Solano, R. Di Candia, Quantum Estimation Methods for Quantum Illumination. *Phys. Rev. Lett.* **118**, 070803 (2017).

29. K.-Y. Spasibko, D.-A. Kopylov, V.-L. Krutyanskiy, T.-V. Murzina, G. Leuchs, M.-V. Chekhova, Multiphoton Effects Enhanced due to Ultrafast Photon-Number Fluctuations. *Phys. Rev. Lett.* **119**, 223603 (2017).

30. L. Zhang, Y. Lu, D. Zhou, H. Zhang, L. Li, G. Zhang, Superbunching effect of classical light with a digitally designed spatially phase-correlated wave front. *Phys. Rev. A.* **99**, 063827 (2019).

31. C.-B. Qin et al., https://arxiv.org/abs/2409.05419 (2024).

32. Z.-Y. Ye, H.-B. Wang, J. Xiong, K.-G. Wang, Antibunching and superbunching photon correlations in pseudo-natural light. *Photonics. Res.* **10**, 668-677 (2022).

33. Y. Zhou, S. Luo, Z.-H. Tang, H.-B. Zheng, H. Chen, J.-B. Liu, F.-l. Li, Z. Xu, Experimental observation of three-photon superbunching with classical light in a linear system. *J. Opt. Soc. Am. B.* **36**, 96-100 (2019).

34. C. Okoth, A. Cavanna, T. Santiago Cruz, M.-V. Chekhova, Microscale Generation of Entangled Photons without Momentum Conservation. *Phys. Rev. Lett.* **123**, 263602 (2019).

35. B. Srivathsan, G.-K. Gulati, B. Chng, G. Maslennikov, D. Matsukevich, C. Kurtsiefer, Narrowband source of transform-limited photon pairs via four-wave mixing in a cold atomic ensemble. *Phys. Rev. Lett.* **111**, 123602 (2013).

36. D. Bhatti, J.-V. Zanthier, G.-S. Agarwal, Superbunching and Nonclassicality as new Hallmarks of Superradiance. *Sci. rep.* **5**, 17335 (2015).

37. F. Jahnke, C. Gies, M. Aßmann, M. Bayer, H.-A.-M. Leymann, A. Foerster, J. Wiersig, C. Schneider, M. Kamp, S. Höfling, Giant photon bunching, superradiant pulse emission and excitation trapping in quantum-dot nanolasers. *Nat. Commun.* **7**, 1-7 (2016).

38. G. Hönig, G. Callsen, A. Schliwa, S. Kalinowski, C. Kindel, S. Kako, Y.

Arakawa, D. Bimberg, A. Hoffmann, Manifestation of unconventional biexciton states in quantum dots. *Nat. Commun.* **5**, 5721 (2014).

39. Z.-Y. Wang, A. Rasmita, G.-K. Long, D.-S. Chen, C.-S. Zhang, O.-G. Garcia, H.-B. Cai, Q.-H. Xiong, W. b. Gao, Optically Driven Giant Superbunching from a Single Perovskite Quantum Dot. *Adv. Opt. Mater.* **9**, 2100879 (2021).

40. G. Qiang, H.-B. Cai, A. Rasmita, R. He, L. Zhou, S.-H. Ru, X.-B. Cai, X.-G. Liu, W.-B. Gao, Giant Superbunching Emission from Mesoscopic Perovskite Emitter Clusters. *Adv. Opt. Mater.* **12**, 1-7 (2023).

41. S. Fiedler, S. Morozov, L. Iliushyn, S. Boroviks, M. Thomaschewski, J.-F. Wang, T.-J. Booth, N. Stenger, C. Wolff, N.-A. Mortensen, Photon superbunching in cathodoluminescence of excitons in WS2 monolayer. *2D. Mater.* **10**, 021002 (2023).

42. M. Manceau, K.-Y. Spasibko, G. Leuchs, R. Filip, M.-V. Chekhova, Indefinite-Mean Pareto Photon Distribution from Amplified Quantum Noise. *Phys. Rev. Lett.* **123**, 123606 (2019).

43. J.-M. Dudley, F. Dias, M. Erkintalo, G. Genty, Instabilities, breathers and rogue waves in optics. *Nat. Photonics.* **8**, 755-764 (2014).

44. É. Rácz, K. Spasibko, M. Manceau, L. Ruppert, M.-V. Chekhova, R. Filip, Quantitative analysis of the intensity distribution of optical rogue waves. *Commun. Phys.* **7**, 1-9 (2024).

45. C.-C. Gerry, P.-L. Knight, Introductory Quantum Optics. (CAM Press, Cambridge, 2005).

46. X. Qi, S.-P. Chen, Z. H. Li, T. Liu, Y. Ou, N. Wang, J. Hou, High-power visible-enhanced all-fiber supercontinuum generation in a seven-core photonic crystal fiber pumped at 1016 nm. *Opt. Lett.* **43**, 1019-1022 (2018).

47. X. Jiang, N.-Y. Joly, M.-A. Finger, F. Babic, G. K.-L. Wong, J.-C. Travers, P.-S.-J. Russell, Deep-ultraviolet to mid-infrared supercontinuum generated in solid-core ZBLAN photonic crystal fibre. *Nature Photonics* **9**, 133-139 (2015).

48. R.-S. Cheng, Y.-Y. Zhou, S.-H. Wang, M.-H. Shen, T. Taher, H.-X. Tang, A 100-pixel photon-number-resolving detector unveiling photon statistics. *Nat. Photonics.* **17**, 112-119 (2023).

49. N. Liu, J. Su, Y. Liu, J. Li, X. Li, Temporal mode properties of Raman scattering in optical fibers. *Opt. Express.* **29**, 13408-13415 (2021).

50. P. A. Moreau, E. Toninelli, T. Gregory, M.-J. Padgett, Ghost Imaging Using Optical Correlations. *Laser. Photonics. Rev.* **12**, 1 (2017).

51. R.-S. Bennink, S.-J. Bentley, R. W. Boyd, "Two-Photon" coincidence imaging

with a classical source. *Phys. Rev. Lett.* **89**, 113601 (2002).

52. F. Di- Lena, F.-V. Pepe, A. Garuccio, M. D'Angelo, Correlation Plenoptic Imaging: An Overview. *Appl. Sci.* **8**, 1958 (2018).

53. A. McCarthy, G.-G. Taylor, J. Garcia Armenta, B. Korzh, D.-V. Morozov, A.-D. Beyer, R.-M. Briggs, J.-P. Allmaras, B. Bumble, M. Colangelo, D. Zhu, K.-K. Berggren, M.-D. Shaw, R.-H. Hadfield, G.-S. Buller, High-resolution long-distance depth imaging LiDAR with ultra-low timing jitter superconducting nanowire single-photon detectors. *Optica* **12**, 168-177 (2025).

54. P. Li, Q. Li, W. Tang, W. Wang, W. Zhang, B.-E. Little, S.-T. Chu, K.-A. Shore, Y. Qin, Y. Wang, Scalable parallel ultrafast optical random bit generation based on a single chaotic microcomb. *Light. Sci. Appl.* **13**, 637-644 (2024).

55. J. Chen, W. Li, Z. Kang, Z. Lin, S. Zhao, D. Lian, J. He, D. Huang, D. Dai, Y. Shi, Single soliton microcomb combined with optical phased array for parallel FMCW LiDAR. *Nat. Commun.* **16**, 1056 (2025).

**Note S1 The measurement principle of noise-tolerant correlated coincidence imaging based on the coincidence counts from super-correlated light sources with high $g^{(2)}(0)$.**

The coincidence counts per second in correlated coincidence imaging (CCI) utilizing a super-correlated light source with high $g^{(2)}(0)$ can be expressed for the object area as:

$$C_o = (g^{(2)}(0) \cdot N_r N_s + N_r \frac{t_1}{t_2} N_n) t_1, \tag{S1}$$

where $g^{(2)}(0)$ is the second-order correlation function of the light source; $N_r$, $N_s$, and $N_n$ are the reference, signal, and noise photon counts per second, respectively; $t_1$ is the coincidence window width and $t_2$ is the pulse period.

The coincidence counts per second of the background area can be expressed as:

$$C_b = N_r \frac{t_1}{t_2} N_n t_1, \tag{S2}$$

The difference in the coincidence counts per second between the object area and the background area can be expressed as:

$$C_o - C_b = g^{(2)}(0) \cdot N_r N_s t_1, \tag{S3}$$

To evaluate the noise tolerance of CCI based on super-correlated light sources with high $g^{(2)}(0)$ in a strong noise environment. The contrast-to-noise ratio (CNR) of the reconstructed image can be expressed as:

$$\text{CNR} = \frac{|C_o - C_b|}{\sqrt{\sigma_{C_o}^2 + \sigma_{C_b}^2}}, \tag{S4}$$

where $C_o$ and $C_b$ denote the coincidence counts per second of object and background areas, respectively; $\sigma_{C_o}$ and $\sigma_{C_b}$ represent the standard deviations of the object and background areas, can be expressed as:

$$\sigma_{C_o}^2 = \left(g^{(2)}(0)\right)^2 \cdot (N_r N_s)^2 \left[\left(\frac{\sigma_{N_s}}{N_s}\right)^2 + \left(\frac{\sigma_{N_r}}{N_r}\right)^2\right] t_1^2 + (N_r \frac{t_1}{t_2} N_n)^2 \left[\left(\frac{\sigma_{\frac{t_1}{t_2} N_n}}{\frac{t_1}{t_2} N_n}\right)^2 + \left(\frac{\sigma_{N_n}}{N_r}\right)^2\right] t_1^2$$

$$= \left[\left(g^{(2)}(0)\right)^2 \cdot (N_r N_s)^2 \left(\frac{1}{N_s} + \frac{1}{N_r}\right) + (N_r \frac{t_1}{t_2} N_n)^2 \left(\frac{t_2}{t_1 N_n} + \frac{1}{N_r}\right)\right] t_1^2, \tag{S5}$$

$$\sigma_{C_b}^2 = (N_r \frac{t_1}{t_2} N_n)^2 \left(\frac{t_2}{t_1 N_n} + \frac{1}{N_r}\right) t_1^2, \tag{S6}$$

Therefore, the CNR of the reconstructed image in intense noise environment can be expressed as:

$$\text{CNR} = \frac{g^{(2)}(0) N_r N_s}{\sqrt{(g^{(2)}(0))^2 \cdot (N_r N_s)^2 (\frac{1}{N_r} + \frac{1}{N_s}) + 2(N_r \frac{t_1}{t_2} N_n)^2 (\frac{1}{N_r} + \frac{t_2}{t_1 N_n})}}. \tag{S7}$$

For the conventional photon-counting imaging (PCI), the photon counts per second of the object area can be expressed as:

$$I_o = N_s + N_n \tag{S8}$$

The photon counts per second of the background area can be expressed as:

$$I_b = N_n \tag{S9}$$

The difference in the photon counts per second between the object area and the background area can be expressed as:

$$I_o - I_b = N_s \tag{S10}$$

The variance of photon counts in the object and background areas can be expressed as follows:

$$\sigma_{I_o}^2 = N_s + N_n \tag{S11}$$

$$\sigma_{I_b}^2 = N_n \tag{S12}$$

Therefore, the CNR of the reconstructed image for PCI in an intense noise environment can be expressed as:

$$\text{CNR} = \frac{N_s}{\sqrt{N_s + 2N_n}} \tag{S13}$$

**Note S2 Theoretical model for generating super-correlated light sources**

The nonlinear process within the photonic crystal fiber (PCF) is analytically segmented into two distinct stages with the predominant nonlinear effects considered separately to elucidate the underlying mechanisms of super-correlated light generation. In the first stage, a pulsed laser with different mean powers is employed to pump the PCF. This excitation leads to the generation of thermal light with diverse intensities, primarily driven by self-phase modulation, cross-phase modulation, and stimulated Raman scattering, among other nonlinear phenomena. In the second stage, the thermal light generated in the first stage serves as a secondary pump source for the PCF. The super-correlated light is subsequently generated through the process of four-wave mixing (FWM), which is driven by the thermal light pumping the PCF.

The output quantum state of the PCF is $|\psi\rangle$, which can be represented as a superposition of several number states:

$$|\psi\rangle = \sum_n c_n |n\rangle, \tag{S14}$$

where $n$ represents the photon number, and $P(n)=|c_n|^2$ denotes the photon number probability.

In the first stage, a pulsed laser pumps the PCF to generate the thermal light field with photon number distribution of the Bose-Einstein distribution, $P_t(n_t)$ expressed as:

$$P_t(n_t) = \frac{\langle n_t \rangle^{n_t}}{(1 + \langle n_t \rangle)^{n_t+1}}, \tag{S15}$$

where $n_t$ represents the thermal photon number. The probability distribution can be approximated by an exponential distribution with a mean $\langle n_t \rangle$, expressed as:

$$P_t(n_t) = \frac{\exp(-n_t / \langle n_t \rangle)}{\langle n_t \rangle}. \tag{S16}$$

In the second stage, the input operator $(\hat{a}_1, \hat{a}_2)$ and output state operator $(\hat{b}_1)$ of FWM in PCF satisfy the relation $\hat{b}_1 = \cosh(\kappa n_t)\hat{a}_1 + \sinh(\kappa n_t)\hat{a}_2^\dagger$, where the input field is considered a vacuum state, $\kappa$ is the FWM conversion efficiency, and $n = \hat{b}_1^\dagger \hat{b}_1 = \sinh^2(\kappa n_t)$ denotes the photon number in the output field near the threshold. The photon number distribution of the super-correlated field pumped by thermal light can be described as (42):

$$P_s(n) = \frac{\exp(-\dfrac{\operatorname{arcsinh}\sqrt{n}}{\kappa\langle n_t\rangle})}{2\sqrt{n(1+n)}\kappa\langle n_t\rangle}. \tag{S17}$$

Varying the pump power or considering additional noise may significantly alter the dominant nonlinear effects, leading to substantial variations in the composition and photon number probability distribution of the generated light field (hereafter referred to as mixed light, comprising coherent, thermal, and super-correlated light). Assuming the mixed light field includes two distinct types of distributed light fields, each characterized by its respective photon number probability distribution:

$$\begin{aligned}
|\psi_1\rangle &= \sum_M c_M |M\rangle, & P_1(M) &= |c_M|^2, \\
|\psi_2\rangle &= \sum_N c_N |N\rangle, & P_2(N) &= |c_N|^2,
\end{aligned} \tag{S18}$$

where $P_1(M)$ and $P_2(N)$ denote the probability distribution of two input fields, while M and N correspond to the photon number of these fields, respectively. The two quantum states above are mixed in a ratio $\xi$, where the ratio $\xi$ is defined by the average photon numbers of the two states.

For the input number states $|M\rangle_1 |N\rangle_2$, we find the mixed light field using the beam splitter model as:

$$|\Psi_{out}\rangle = \frac{1}{\sqrt{M!N!}}(t\hat{a}_1 - r\hat{a}_2)^M (t\hat{a}_2^\dagger + r\hat{a}_1^\dagger)^N |0\rangle = \sum_{k=0}^{M+N} c_k |k, M+N-k\rangle, \tag{S19}$$

where $t = \sqrt{\xi}$ and $r = \sqrt{1-\xi}$ are the complex amplitude transmissivity and reflectivity of the beam splitter, respectively. The coefficient $c_k$ characterizes the weighting contribution of photon number k in the mixed field. Therefore, for the input states described in Eq. (S5), the photon number probability distribution of the mixed light field can be determined through a superposition of a series of quantum states presented in Eq. (S6), as expressed:

$$P_{mix}(k) = \sum_{M=0}^{\infty} P_1(M) \sum_{N=0}^{\infty} P_2(N) \cdot |c_k|_{MN}^2. \tag{S20}$$

The second-order correlation function can be expressed as:

$$g^{(2)}(\tau) = \frac{\langle :\hat{I}(t)\hat{I}(t+\tau): \rangle}{\langle \hat{I}(t)\rangle \langle \hat{I}(t+\tau)\rangle}, \tag{S21}$$

where $\hat{I}(t)$ and $\hat{I}(t+\tau)$ represent the instantaneous light intensity measured by the two single-photon detectors respectively, the symbol :: represents the normal ordering, $\tau$ represents the time delay, and $\langle \cdot \rangle$ represents the time or ensemble average. Taking a single-mode light field as an example, the detector light intensity operator can be expressed as:

$$\hat{I}(t) = \hat{a}^\dagger \hat{a}, \tag{S22}$$

where, $\hat{a}^\dagger$ is the generation operator and $\hat{a}$ is the annihilation operator.

In the experiment, the second-order correlation function $g^{(2)}(0)$ can be expressed as the ratio of biphoton coincidence counts to accidental coincidence counts. The probability of coincidence count and accidental coincidence count can be obtained respectively:

$$N_2 = \mathrm{Tr}\left\{\hat{a}^\dagger \hat{a}^\dagger \hat{a} \hat{a} \hat{\rho}\right\} = \sum_{n=0}^{\infty} P_n \cdot n(n-1),$$

$$N_1 = \mathrm{Tr}\left\{\hat{a}^\dagger \hat{a} \hat{\rho}\right\} = \sum_{n=0}^{\infty} P_n \cdot n, \tag{S23}$$

where, $\hat{\rho} = \sum_{n=0}^{\infty} P_n |n\rangle\langle n|$ represents the density operator and $P_n$ represents the probability distribution of the mixed field. Further, the zero-delay second-order correlation function can be obtained (44):

$$g^{(2)}(0) = \frac{\sum_n P_n \cdot n(n-1)}{\left(\sum_n P_n \cdot n\right)^2}. \tag{S24}$$

The photon number statistics of coherent light follow a Poisson distribution and can be expressed as:

$$P_c(n) = \frac{\langle n \rangle^n e^{-\langle n \rangle}}{n!}, \tag{S25}$$

where $n$ and $\langle n \rangle$ are photon numbers and mean photon numbers per pulse. Thermal light satisfies the Bose-Einstein distribution and can be expressed as:

$$P_t(n) = \frac{\langle n \rangle^n}{(1+\langle n \rangle)^{n+1}}. \tag{S26}$$

Figure. S1A illustrates the theoretical predictions of the photon number probability distribution (as derived from Eq. (S17)) for the super-correlated light at different mean photon numbers $\langle n \rangle$. Generally, the photon number probability distributions of coherent (CL) and thermal light (TL) sources follow the Poisson distribution and the Bose-Einstein distribution, respectively (Eqs. (6) and (7)). In contrast, the super-correlated light manifests a fundamentally different statistical behavior, displaying a significantly broader photon number distribution. For instance, when $\langle n \rangle = 1.3 \times 10^{-1}$ the probability of detecting 16 photons is enhanced by 10 and 23 orders of magnitude relative to thermal light and coherent light, respectively. We further simulated the second-order correlation function $g^{(2)}(0)$ of super-correlated light using Eq. (S24). The results are represented by the orange solid line in Fig. S1B. As $\langle n \rangle$ decreases, the intensity correlation increases gradually, demonstrating significantly higher values compared to both thermal light (blue dotted line, $g^{(2)}(0) = 2$) and coherent light (red dotted line, $g^{(2)}(0) = 1$). Considering both coherent and thermal noise contributions, we observe a significant change in intensity correlation for mixed light as a function of the super-correlated light mean photon number $\langle n \rangle$, as shown in Fig. S1B. Notably, for a fixed noise photon level, correlation strength first increases, peaks, then decreases with decreasing photon number. It eventually reaches distinct maxima in each case as the relative weight of correlated photons becomes increasingly dominant. The extremely high multiphoton emission probability of the super-correlated light source significantly enhances the probability of detecting correlated photons, making it substantially higher than the accidental coincidence probability caused by noise photons. These combined properties make it more suitable for single-photon-level correlated imaging applications in extreme environments.

## Note S3 Numerical simulation of PCF

The PCF is a single-core fiber consisting of five layers of air holes. The core diameter ($d$) is 4.6 µm, with air hole diameters ($\Lambda$) and spacings ($d_0$) of 1.9 µm and 3.3 µm in the cladding, respectively. According to previous works, the generation of spectra and the occurrence of nonlinear effects are influenced by the dispersion and the nonlinear coefficient of the PCF. Employing the finite element method, we numerically simulated both the dispersion characteristics and the nonlinear coefficient of our PCF. The simulation results indicate that the zero-dispersion wavelength (ZDW) is 1030 nm, and the nonlinear coefficient at 1064 nm is 11.03 $W^{-1}$/km (Fig. S3).

## Note S4 Variation of optical field properties with different proportions of attenuations under the same pump power

We utilized the HBT system to verify the relationship between the second-order correlation function $g^{(2)}(0)$ and the mean photon number $\langle n \rangle$ under the same pump power. This system is equipped with two channels, which we define as CH1 and CH2. We used neutral density filters as precision optical attenuators to precisely control the mean photon number. These filters demonstrate well-defined linear transmission characteristics with wavelength-independent attenuation properties across the operational spectral range. At the same pump power, the linear attenuation ratios of the reference path and detection path in the HBT system are nearly equivalent (Fig. S4A). Super-correlated light sources demonstrated exceptional photon correlation properties. Intriguingly, although the mean photon number per pulse $\langle n \rangle$ decreases, a counterintuitive increase in the second-order correlation function $g^{(2)}(0)$ is observed, revealing the non-classical nature of the photon statistics. This phenomenon originates from the prevalence of multi-photon pulses in super-correlated light which maintain a high likelihood of correlated photon events. As the attenuation ratio increases, photon sparsity in the time domain becomes more pronounced. Although the coincidence counts per second of the super-correlated light decline, the rate of reduction is substantially lower than the probability of accidental coincidence counts caused by the coincidence counts reference and detection paths with different time pulses (Fig. S4B), resulting in a notable increase in $g^{(2)}(0)$. For comparison, we measured the coincidence counts, accidental coincidence counts and $g^{(2)}(0)$ of coherent light generated by a 1550 nm mode-locked pulsed laser. The output power of the pulsed laser is maintained constant, with photon count attenuation achieved through a neutral density filter. Photon counts per second in both the reference path and detection path of the HBT system exhibit linear attenuation (Fig. S4C). However, the coincidence and accidental coincidence counts per second remain consistent (Fig. S4D), resulting in a constant value of $g^{(2)}(0)$=1. This exceptional characteristic of super-correlated light highlights its elevated multi-photon probability, forming the foundation for attaining extraordinarily high $g^{(2)}(0)$ values.

**Note S5 Measurement of the photon number probability distribution and the second-order correlation function $g^{(2)}(0)$ of coherent light**

To evaluate the reliability of the system, we characterize the photon number probability distribution and the second-order correlation function $g^{(2)}(0)$ for coherent states with different mean photon numbers using a mode-locked pulsed laser (Rainbow 1550 OEM, NPI LASERS) operating at a central wavelength of 1550 nm, an output power of 100 mW, and a repetition frequency of 20 MHz. The mean photon number $\langle n \rangle$ is varied through the use of multiple neutral density filters. SNSPD and TCSPC are employed to measure the photon number probability distribution and the second-order correlation function $g^{(2)}(0)$ under varying mean photon numbers $\langle n \rangle$. Figure S5B illustrates the measured and simulated results for coherent states with various mean photon numbers. For mean photon numbers of $2.5 \times 10^{-1}$, $2.7 \times 10^{-2}$, and $2.9 \times 10^{-3}$, respectively, the measured photon number probability distributions (represented by bar charts) follow the Poisson distribution (represented by dot plot). Figure S5C displays a typical second-order correlation function $g^{(2)}(0)$ for a mean photon number of $2.5 \times 10^{-1}$, with $g^{(2)}(0)$=1.03±0.06. Figure S5D shows $g^{(2)}(0)$ as a function of the mean photon number. As expected, $g^{(2)}(0)$ remains approximately unity for coherent light.

**Note S6 Generation and measurement of the photon number probability distribution and the second-order correlation function $g^{(2)}(0)$ of thermal light**

The 1550 nm mode-locked (Rainbow 1550 OEM, NPI LASERS) laser was amplified by an erbium-doped fiber amplifier (EDFA) to generate a pump source for a 500 m single-mode fiber (SMF-28e, Corning) During the spectral broadening process, Raman peaks were obtained by using DWDM (MCDWDM-1-C15-00-S2-08-L-FA, DK Photonics) filtering with a central wavelength of 1565 nm and a spectral width of 0.8 nm to simulate thermal light (Fig. S5A). The intensity correlation function was measured using an identical experimental approach to that employed for coherent light. Figure S5E illustrates the measured and simulated photon number probability distributions for thermal light at different mean photon numbers $\langle n \rangle$. When the mean photon numbers are $1.7 \times 10^{-1}$, $1.8 \times 10^{-2}$, and $1.5 \times 10^{-3}$, respectively, the measured photon number probability distribution (bar chart) matches the Bose-Einstein distribution (dot plot) from the theoretical simulation. Figure S5F shows a typical second-order correlation function $g^{(2)}(0)$ for a mean photon number of $1.7 \times 10^{-1}$ with $g^{(2)}(0)$=1.85±0.1. Figure S5G presents $g^{(2)}(0)$ as a function of the mean photon number $\langle n \rangle$. As anticipated, the $g^{(2)}(0)$ value remained basically constant for the thermal light.

**Note S7 Noise-tolerant correlated imaging based on super-correlated light sources for different measurement times**

Most single-photon imaging schemes based on entangled photons utilize electron-multiplying charge-coupled devices (EMCCD) cameras to measure biphoton coincidence counts. However, due to the low frame rate, the imaging time can extend to tens of hours. Here, we verified the image quality of the super-correlated light source at the average integration time per pixel (100 ms, 300 ms, 500 ms). In the absence of external noise, although the PSNR and CNR decrease with shorter integration times, the imaging quality remains relatively high. These results demonstrate that the super-correlated light source exhibits strong robustness across different imaging durations (Fig. S8).

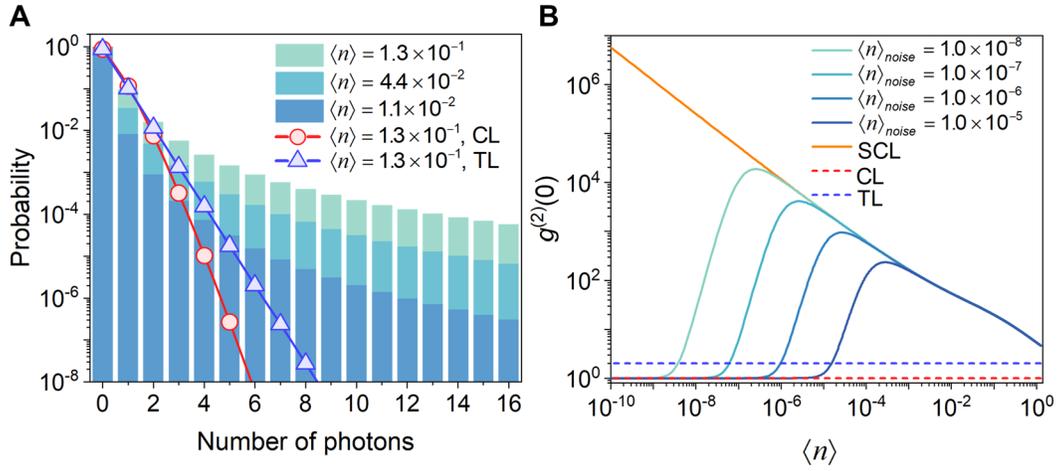

**Fig. S1 Theoretical simulation of photon number probability distribution and second-order correlation function $g^{(2)}(0)$.**

(A) Theoretical photon number probability distributions for super-correlated light (SCL), coherent light (CL), and thermal light (TL) with varying mean photon numbers per pulse $\langle n \rangle$. (B) Theoretical predictions of $g^{(2)}(0)$ for the SCL as a function of $\langle n \rangle$ at different levels of noisy photons $\langle n \rangle_{noise}$.

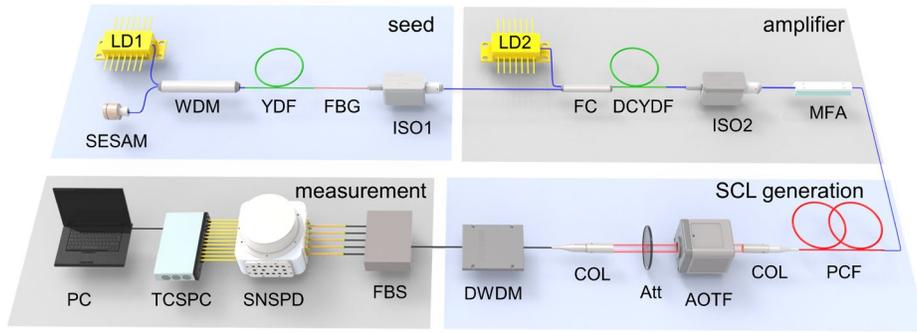

**Fig. S2 Schematic diagram of the super-correlated light (SCL) generation and measurement setup.**
LD1 and LD2, semiconductor laser; SESAM, semiconductor saturable absorption mirror; WDM, wavelength division multiplexer; YDF, ytterbium-doped fiber; FBG, fiber Bragg grating; ISO, fiber isolator; FC, fiber combiner; DCYDF, double-clad ytterbium-doped fiber; MFA, mode field adapter; PCF, photonic crystal fiber; Att, neutral density filters; COL, fiber optic collimator; AOTF, acousto-optic tunable filter; DWDM, dense wavelength division multiplexer; FBS, fiber beam splitter; SNSPD, superconducting nanowire single photon detector; TCSPC, time-correlated single photon counting; PC, computer.

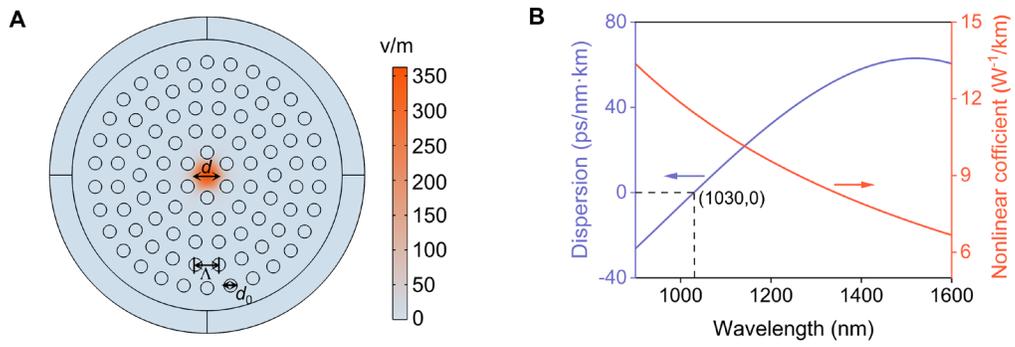

**Fig. S3 Numerical simulation results of PCF.**
(**A**) Electric field distribution of the transmission mode in the PCF. $d$ = 4.6 μm, $d_0$ = 3.3 μm, $\Lambda$ =1.9 μm. (**B**) Dispersion and nonlinear coefficient curves of the PCF.

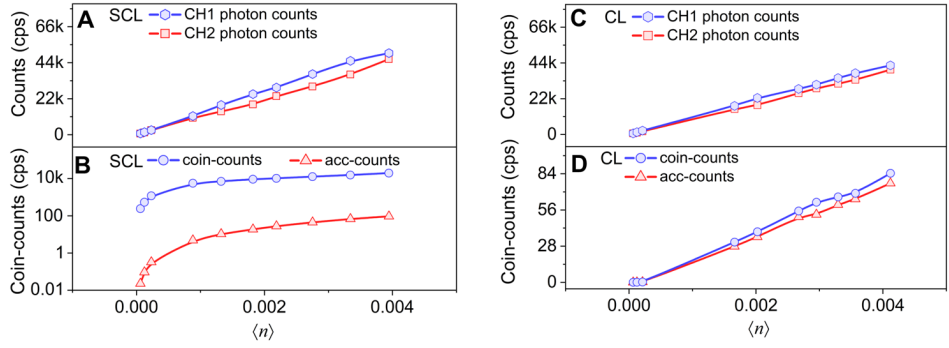

**Fig. S4 For different ⟨$n$⟩, Comparison of the coincidence photon counts and photon counts per second between super-correlated light and coherent light.**
Photon counts per second for the CH1 and CH2 in the HBT system for super-correlated light (**A**) and coherent light (**C**). Coincidence counts (coin-counts) and accidental coincidence counts (acc-counts) per second for super-correlated light (**B**) and coherent light (**D**).

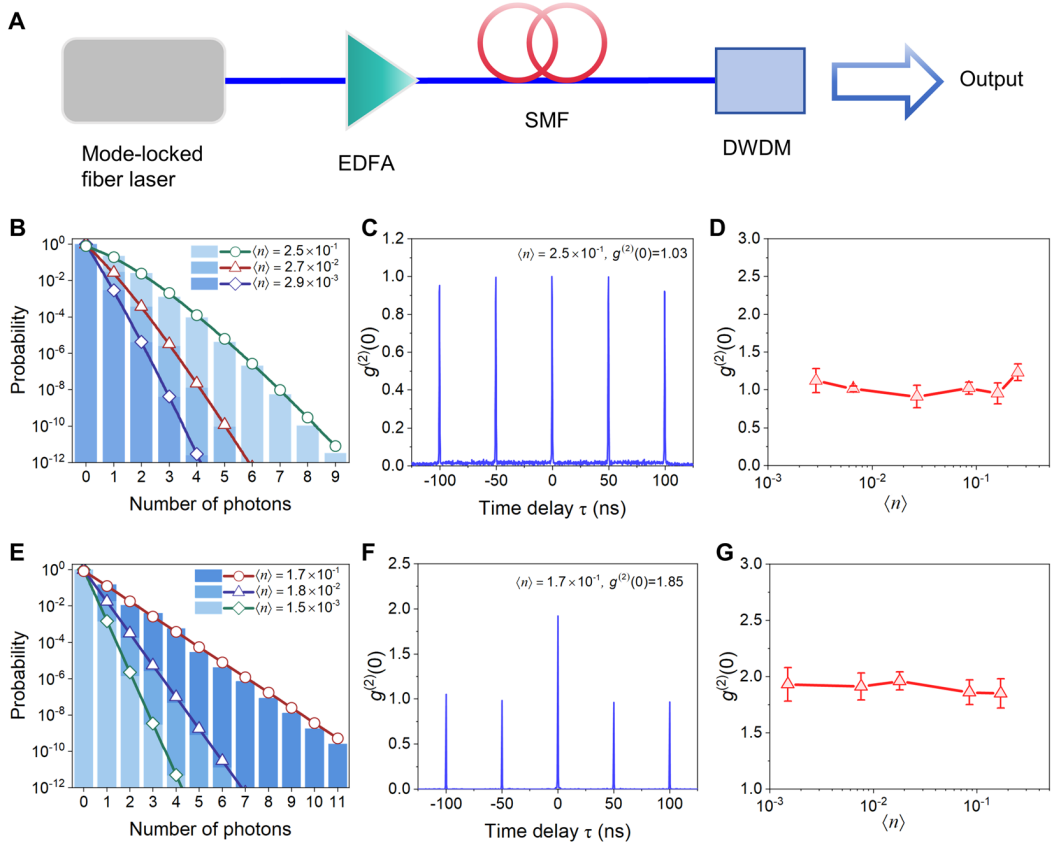

**Fig. S5 Second-order correlation function $g^{(2)}(0)$ and photon number probability distribution of coherent light and thermal light.**
(**A**) The generating device of the thermal light source. (**B**), (**E**) Photon number probability distributions for coherent and thermal light, showing experimental measurements (bar chart) and theoretical simulations (dot plot) for different mean photon numbers $\langle n \rangle$. (**C**), (**F**) The second-order correlation function $g^{(2)}(\tau)$ as a function of the time delay $\tau$, for coherent and thermal light, respectively. (**D**), (**G**) Measured $g^{(2)}(0)$ as a function of the mean photon number per pulse for coherent and thermal light, respectively.

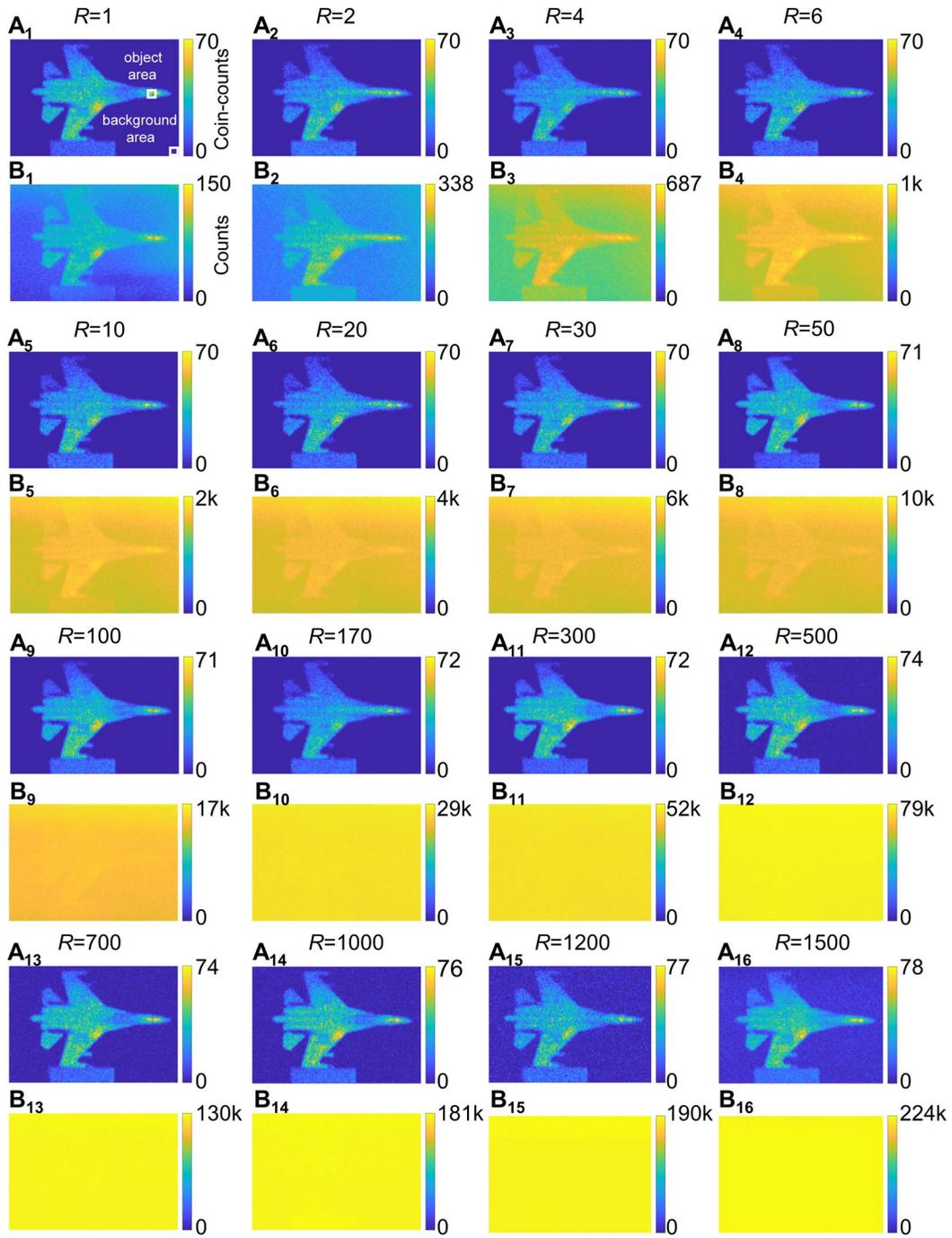

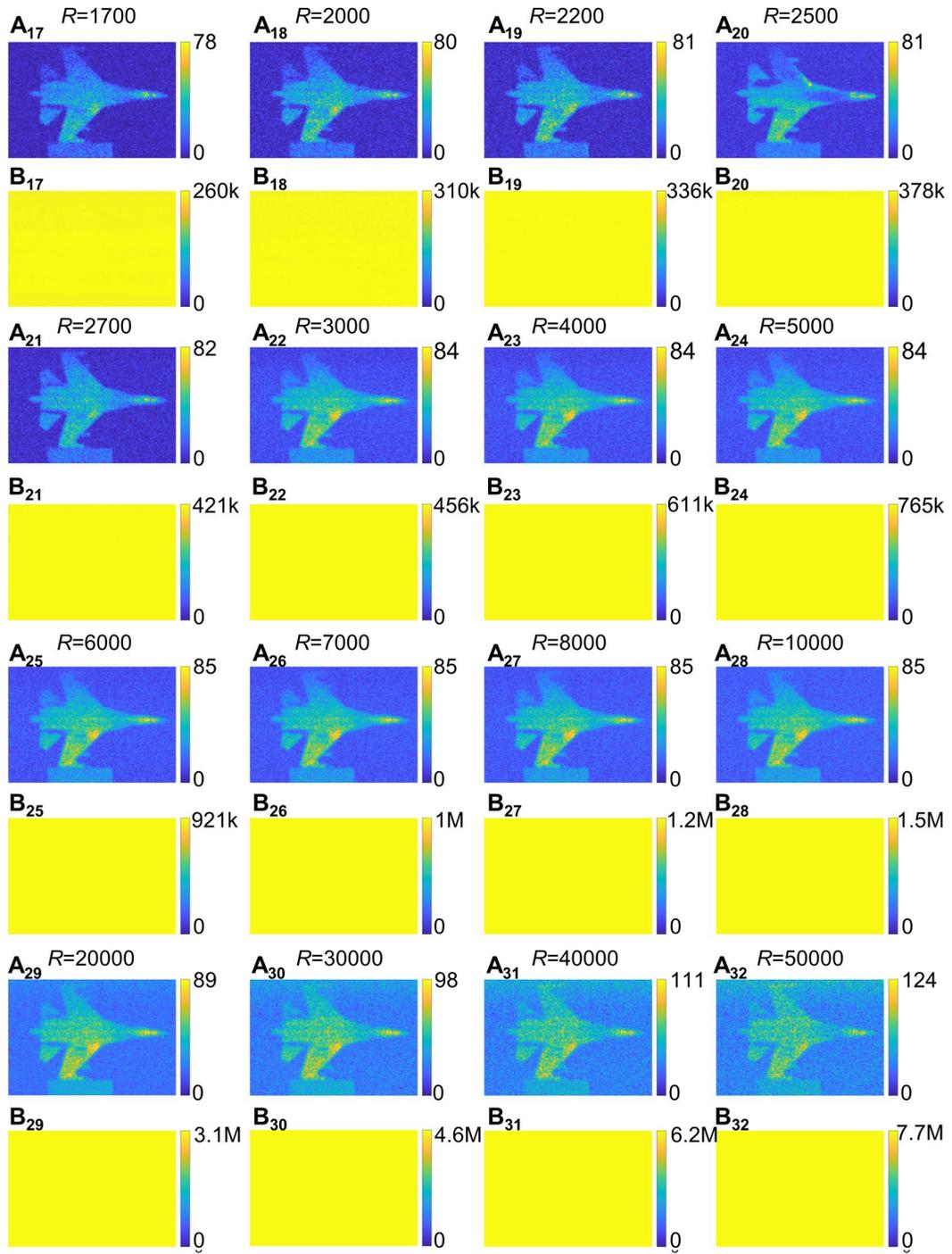

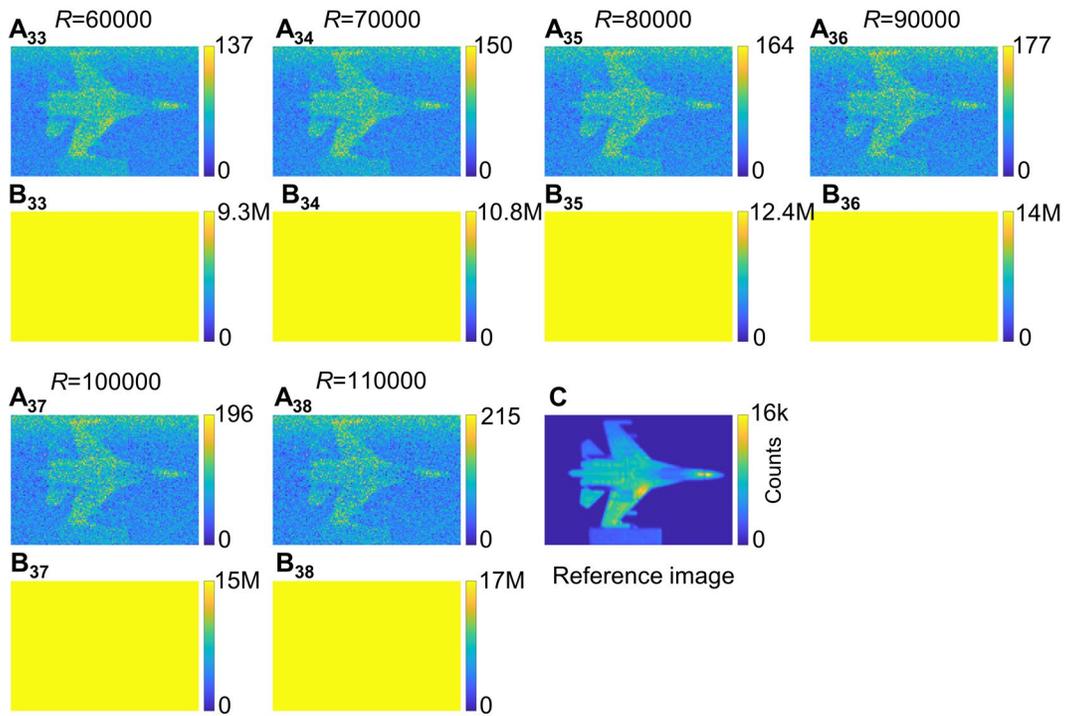

**Fig. S6 Image reconstruction results based on CCI and PCI under different noise conditions.** (**A₁-A₃₈**) Imaging results of CCI. The color bar indicates the coincidence count values (coin-counts). (**B₁-B₃₈**) Imaging results of PCI. The color bar indicates the photon count values. (**C**) Illuminate the target object with a high photon number and a long measurement time as the reference image. The color bar indicates the photon count values.

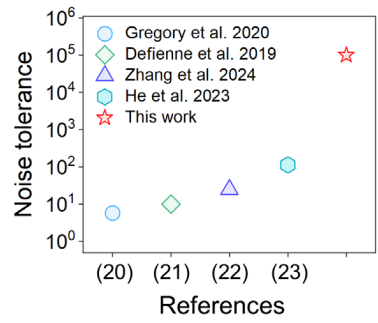

**Fig. S7 Comparison of the noise tolerance performance of correlated imaging by different approaches.**

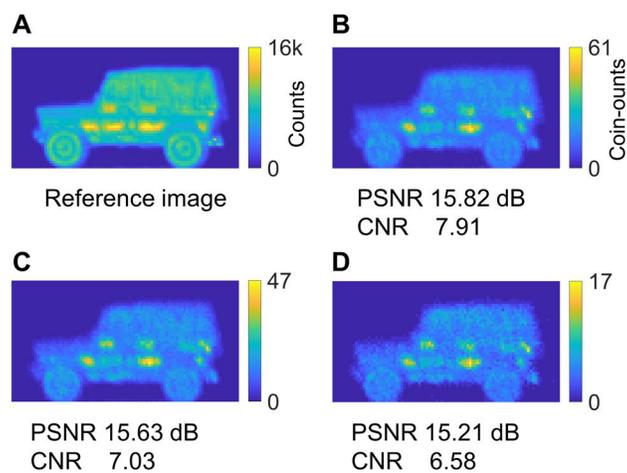

**Fig. S8 Imaging results of CCI under different measurement times for per pixel.**
(A) Illuminate the target object with a high photon number and a long measurement time as the reference image. The color bar indicates the photon count values. The measurement time per pixel is (B) 500 ms, (C) 300 ms, and (D) 100 ms. The color bar indicates the coincidence count values (coin-counts).